# Anatomy of Type-X Spin-Orbit Torque Switching

Yan-Ting Liu[1†], Chao-Chung Huang[1†], Kuan-Hao Chen[1], Yu-Hao Huang[1], Chia-Chin Tsai[1], Ting-Yu Chang[1] and Chi-Feng Pai[1,2*]

[1]*Department of Materials Science and Engineering, National Taiwan University, Taipei 10617, Taiwan*

[2]*Center of Atomic Initiative for New Materials, National Taiwan University, Taipei 10617, Taiwan*

Using type-*x* spin-orbit torque (SOT) switching scheme, in which the easy axis (EA) of the ferromagnetic (FM) layer and the charge current flow direction are collinear, is possible to realize a lower-power-consumption, higher-density, and better-performance SOT magnetoresistive random access memory (SOT-MRAM) as compared to the conventional type-*y* design. Here, we systematically investigate type-*x* SOT switching properties by both macrospin and micromagnetic simulations. The out-of-plane external field and anisotropy field dependence of the switching current density ($J_\text{sw}$) is first examined in the ideal type-*x* configuration. Next, we study the FM layer canting angle ($\varphi_\text{EA}$) dependence of $J_\text{sw}$ through macrospin simulations and experiments, which show a transformation of switching dynamics from type-*x* to type-*y* with increasing $\varphi_\text{EA}$. By further integrating field-like torque (FLT) into the simulated system, we find that a positive FLT can assist type-*x* SOT switching while a negative one brings about complex dynamics. More crucially, with the existence of a sizable FLT, type-*x* switching mode results in a lower critical switching current than type-*y* at current pulse width less than ~ 10 ns, indicating the advantage of employing type-*x* design for ultrafast switching using materials systems with FLT. Our work provides a thorough examination of type-*x* SOT scheme with various device/materials parameters, which can be informative for designing next-generation SOT-MRAM.

[†] Y.-T. L. and C.-C. H. contributed equally to this work
[*] Email: cfpai@ntu.edu.tw

# I. INTRODUCTION

Current-induced bulk spin Hall effect [1,2] (SHE) from heavy metal (HM) and interfacial Rashba Edelstein effect [3] (REE) are capable of exerting spin-orbit torques (SOTs) upon an adjacent ferromagnetic (FM) layer, providing a promising approach to manipulate the magnetization to achieve high frequency oscillation [4-6], domain wall motion [7-9], and SOT-induced magnetization switching [3,10-14]. Spin torques (such as spin-transfer torque (STT) and SOT) driven magnetization switching is a promising technology for magnetic random-access memory (MRAM) applications. Unlike the vulnerability of conventional STT-MRAM based on STT mechanism [15], SOT-MRAM has the advantage of low power-consumption, faster switching and separation of read and write paths to guarantee high endurance, approaching industrial needs. There are three different types of SOT switching schemes, *i.e.*, type-*x*, type-*y* and type-*z* [16-18]. Type-*z* geometry has a FM pillar with perpendicular magnetic anisotropy (PMA). For type-*x* (type-*y*) geometry, the magnetic easy axis of the FM pillar lies in the FM film plane and is collinear with (orthogonal to) the applied current. It has been shown that type-*y* requires current with a longer pulse width to achieve deterministic switching, which limits the performance of type-*y* device in the short pulse regime [18,19]. In contrast, for type-*x* and type-*z* geometries, where the orthogonal spin polarization ($\hat{\sigma}$) dominates, the SOT-induced magnetization switching takes place with little precession, making them less sensitive to the applied current pulse width. At the short current pulse regime, compared to type-*y* geometry, smaller switching current density ($J_{sw}$) of type-*x* and type-*z* has been demonstrated via numerical simulation [16] and experiment [19]. However, due to the non-collinearity between $\hat{\sigma}$ and magnetization in type-*z* and type-*x* configurations, an external field is required to break symmetry and achieve deterministic switching [20]. For type-*z*, field-free deterministic switching can be realized by introducing a voltage-controlled electric field [21], the assistance of exchange bias field from the antiferromagnetic layer [22], wedge structure [12,23], insertion of an additional FM layer with in-plane anisotropy [24-26], the geometrical domain-wall pinning [27], the presence of competing spins from materials with opposite spin Hall angles [28,29], or lateral SOT from localized laser annealing [30]. These approaches are typically challenging to achieve in nano-sized devices. For type-$x$, there is a much simpler way to achieve field-free switching through structural engineering: introducing a small tilting angle between the easy-axis of the FM pillar and the current channel [16,19,31]. A brief summary of the characteristics of these three SOT switching schemes can be found in Table. 1.

In this work, we systemically investigate type-$x$ SOT switching properties in a nano-sized HM/FM bilayer structure by macrospin and micromagnetic simulations. First, we demonstrate that $J_{sw}$ in type-$x$ SOT switching scheme can be modified by external field $\mu_0 H_z$, uniaxial anisotropy field and shape anisotropy field. Next, field-free switching for type-$x$ geometry is verified by canting the major axis orientation of the FM pillar. Through canting angle dependence simulations, we observe a sharp transition from type-$x$ to type-$y$ switching behavior as the canting angle increases. We also experimentally verify this switching dynamic change by measuring $J_{sw}$ vs. canting angle $\varphi$ in representative HM/FM bilayer devices (HM = W; FM = CoFeB). From pulse width dependence simulation using type-$x$ and type-$y$ geometries, we show that the switching current ($I_{sw}$) of type-$x$ is indeed smaller than that of type-$y$, but only in the *extremely short* pulse width (< 0.2 ns) regime. A detailed switching phase diagram with various field-like torque (FLT) to damping-like torque (DLT) ratios and applied current densities ($J_{applied}$) is obtained to gain insight into all possible magnetization dynamics. For FLT/DLT > 0, only deterministic switching and non-switching can be obtained. $J_{sw}$ reduces significantly with increasing FLT. On the other hand, for FLT/DLT < 0, more complex magnetization dynamics can emerge. Lastly, through both macrospin and micromagnetic simulations, we show that $I_{sw}$ of the field-free type-$x$ scheme can be reduced by increasing FLT/DLT ratio. The crossover of $I_{sw}$ vs. $t_{pulse}$ curves for type-$x$ and type-$y$ gradually moves to a longer pulse regime (> 1 ns) with increasing FLT/DLT ratio, which points out the benefit of employing magnetic heterostructures with sizable FLT for type-$x$ SOT applications.

## II. RESULTS AND DISSCUSION

The magnetization dynamics in ferromagnetic (FM) layer is described by the Landau-lifshitz-Gilbert (LLG) equation with an additional damping-like SOT (DLT) term:

$$\frac{d\mathbf{m}}{dt} = -\gamma\,\mathbf{m}\times\mathbf{H}_{\mathrm{eff}} + \alpha\,\mathbf{m}\times\frac{d\mathbf{m}}{dt} + \gamma\frac{\hbar\xi_{\mathrm{DL}}J_{\mathrm{applied}}}{2eM_{\mathrm{s}}t_{\mathrm{FM}}}\mathbf{m}\times(\boldsymbol{\sigma}\times\mathbf{m}), \qquad (1)$$

where $\gamma$ is the gyromagnetic ratio, $\alpha$ the Gilbert damping constant, $M_s$ the saturation magnetization of the FM pillar, $J_{\mathrm{applied}}$ the applied current density, $t_{\mathrm{FM}}$ the thickness of the FM layer, $\boldsymbol{\sigma}$ the orientation of spin polarization accumulated at the HM/FM interface, and $\xi_{\mathrm{DL}}$ the damping like (DL) SOT efficiency. The effective field $\mathbf{H}_{\mathrm{eff}} = \mathbf{H}_{\mathrm{ext}} + \mathbf{H}_{\mathrm{K}} + \mathbf{H}_{\mathrm{d}}$ is composed of the external field, uniaxial anisotropy field and shape anisotropy field. To get closer to realistic scenarios, we use the parameters listed in Table 2. The size of the FM nanopillar is $60 \times 30 \times 1$ nm$^3$ (L × W × t$_{\mathrm{FM}}$). Since the experimentally determined magnitude of $\xi_{\mathrm{DL}}$ for tungsten (W) is ~ 0.3 to 0.5 [11,32-35], we tentatively set $|\xi_{\mathrm{DL}}| = 0.5$ for the HM layer in our simulation. The width of HM layer is equal to the width of the FM pillar along $y$. The thickness of the HM layer ($t_{\mathrm{HM}}$) is 3 nm, which is longer than the spin diffusion length of W [13,36].

## a. Three schemes of SOT switching

Schematics of type-*x*, *y*, *z* switching configurations are shown in Fig. 1 (a) to (c). For type-*x* (type-*y*) geometry, the easy axis of the FM pillar is collinear with (orthogonal to) the current channel direction. Note that the easy axis of the FM pillar is mainly induced by shape anisotropy for in-plane magnetized systems. For type-*z* geometry, the easy axis is perpendicular to the FM pillar plane (*xy* plane). The current is applied along ±*x* direction and σ̂ at HM/FM interface thus points at ±*y* direction. To break symmetry, we set the external field $\mu_0H_x$ = +20 mT and $\mu_0H_z$ = +20 mT for type-*z* and type-*x* configuration, respectively. Fig. 1(d) to (f) show the time evolution of magnetization for these three schemes with a 5 ns duration of the applied current (and external field) and a 5 ns relaxation. We also show the 3D plot of magnetization trajectories with ($m_x$, $m_y$, $m_z$) coordinate in Fig. 1(g) to (i).

For type-*y* geometry where collinear spin polarization dominates, the switching dynamics are quite different from those of type-*x* and type-*z*. Type-*y* system goes through more precessions during the SOT-driven magnetization switching process. For the systems where orthogonal spin polarization dominates, the magnetization of type-*x* (type-*z*) is almost directly switched from +*x* (+*z*) to the -*x* (-*z*) region without precession when the current pulse is turned on. Since the magnetization for type-*x* geometry is affected by two different shape anisotropy fields ($\mathbf{H}_{d,y}$ and $\mathbf{H}_{d,z}$ +$H_K$) along *y* and *z* directions while type-*z* geometry is only influenced by uniaxial anisotropy, the two switching modes are similar but with a slight variation.

## b. $\mu_0H_z$ dependence of type-*x* SOT switching

We first examine the $\mu_0H_z$ dependence of $J_{sw}$ in the type-*x* case. The representative device geometry is set as 60 × 30 × 1 nm³ and $K_u$ is set to be -8 × 10⁵ J/m³ with an uniaxial anisotropy unit vector $u$ = [0, 0, 1]. An external field $\mu_0 H_z$ for the type-$x$ structure is required to achieve deterministic switching. $J_{sw}$ is defined as the threshold current density value at which magnetization can be deterministically switched. Initial position of the magnetization is defined as $m_x$ ~ +1. As shown in Fig. 2(a), $J_{sw}$ decreases linearly to $\mu_0 H_z$. Since $\sigma$ is perpendicular to the easy axis direction in both type-$z$ and type-$x$ schemes, the switching dynamics of type-$x$ should be similar to that of type-$z$. We then fit the data by the equation derived analytically by K.S. Lee *et al*. [37] (modified for type-$x$):

$$\frac{dJ_{sw}}{d\mu_0 H_z} = -\frac{2e}{\sqrt{2}\hbar}\frac{M_s t_{FM}}{\xi_{DL}}, \tag{2}$$

where $dJ_{sw}/d\mu_0 H_z$ is the extracted slope from the $J_{sw}$ vs. $\mu_0 H_z$ curve. From the linear fitting, we can obtain the slope $dJ_{sw}/d\mu_0 H_z$= -5.73×10¹² A/Tm² and the $x$-intercept $\mu_0 H_{z,0}$ = 0.40 T. The magnitude of the slope ($dJ_{sw}/d\mu_0 H_z$) is comparable to the number extracted for type-$z$ system [37]. More importantly, S. Fukami's seminal work on type-$x$ switching also demonstrated that the linear trend between switching current density and $\mu_0 H_z$ can be observed via both experiments and simulations [16]. Therefore, we find that the external bias field dependence of SOT-driven switching current densities for type-$x$ (using $\mu_0 H_z$) and type-$z$ schemes (using $\mu_0 H_x$) are quite similar.

### c. Uniaxial and shape anisotropy dependence of type-$x$ switching

Since $J_{sw}$ also depends on the interfacial, bulk, and shape anisotropy of the device,[16] we examine the influence of uniaxial anisotropy energy density ($K_u$) and the shape effect (aspect ratio) of the FM pillar on $J_{sw}$ for type-$x$ geometry. As shown in Fig. 2(b), when $K_u$ varies from 0 to -1× $10^6$ J/m$^3$ with $u$ = [0, 0, 1], both the effective out-of-plane anisotropy field $\mu_0H_{k,out}$ and $J_{sw}$ ($\mu_0H_z$ = +20 mT) increase with respect to the magnitude of $K_u$. Note that $K_u$ = 0 means that only shape anisotropy is considered. We set $K_u$ as -8 × $10^5$ J/m$^3$ for further simulations.

To investigate the importance of shape engineering, we calculate $J_{sw}$ as functions of pillar aspect ratio, minor axis length and thickness of the FM layer. The shape anisotropy field is estimated by - $N_dM_S$, where $N_d$ is the demagnetization factor and is calculated from major axis length, minor axis length and thickness of the FM pillar [38]. In the aspect ratio dependence simulations, the minor axis length and thickness of the FM pillar are fixed at 30 nm and 1 nm, respectively. We modify the length of the pillar major axis from 30 nm to 218 nm (aspect ratio = 1.00 to 7.25). As shown in Fig. 2(c), $J_{sw}$ becomes larger with increasing aspect ratio of the FM pillar and saturates at aspect ratio ≈ 7.0. Out-of-plane and in-plane effective anisotropy fields, $\mu_0H_{k,out}$ and $\mu_0H_{k,in}$, are also shown in Fig. 2(d) with various aspect ratios. $\mu_0H_{k,out}$ only slightly increases with increasing aspect ratio, while $\mu_0H_{k,in}$ enhances significantly with aspect ratio, which suggests that the aspect ratio affects $\mu_0H_{k,in}$ and then $J_{sw}$. $J_{sw}$ vs. minor axis length and $J_{sw}$ vs. FM layer thickness also mainly depend on $\mu_0H_{k,in}$ vs. minor axis length or FM thickness (details of minor axis length and FM pillar thickness dependence can be found in Supplementary Materials S1[39]).

According to the results of shape-dependent simulations, we conclude that $J_{sw}$ strongly depends on $\mu_0H_{k,in}$, which can be modified by the shape of the FM pillar. More specifically, a lower $J_{sw}$ can be achieved by reducing $\mu_0H_{k,in}$ through shape engineering by choosing a smaller aspect ratio. However, since the thermal stability of the device is proportional to the volume of the FM pillar and $\mu_0H_{k,in}$ (thereby its aspect ratio), an optimized aspect ratio should be chosen in realistic memory applications in order to lower the switching current while maintaining reasonable thermal stability. For simplicity and generality, we choose the major axis length, minor axis length, and thickness of the FM pillar as 60 nm, 30 nm, and 1 nm, respectively, for the following simulation tests.

**d. Pulse width dependence of type-*x* and type-*y* switching**

To compare the performance between type-*x* and type-*y* structures at short pulse width regime, we calculate $J_{sw}$ with various pulse widths (0.1 ns $\leq t_{pulse} \leq$ 10 ns) followed by a 5 ns of relaxation. It is noted that the pillar of type-*x* system is canted toward -*y* direction about 1 deg to break symmetry, thus no $\mu_0 H_z$ is required [16,19,31]. The geometries of the simulated devices, type-*x* ($\varphi$ = 1 deg) and type-*y* ($\varphi$ = 90 deg), are shown in Fig. 3(a). The current channel width is set to be the same as the pillar size along *y* direction, *i.e.*, 30 nm for type-*x* and 60 nm for type-*y*. It can be seen that one of the advantages of employing type-*x* structure is the possibility of using a much narrower current channel. Based on these assumptions, we calculate the critical switching current $I_{sw}$ for both type-*x* and type-*y* structures with various pulse widths, which is shown in Fig. 3(b). The results indicate $I_{sw}$ vs. $t_{pulse}$ of type-*x* changes more gradually than type-*y*. The two sets of data cross over at $t_{pulse} \sim$ 0.23 ns, which is denoted as $t_{CO}$. This suggests that only in the extremely short pulse regime (sub-ns), type-*x* outperforms type-*y* switching.

For fast precessional switching in the ns or sub-ns regimes, the pulse width dependence of $I_{sw}$ can be further fitted by [15,40-42]:

$$I_{sw} = I_{sw,0}\left(1 + \frac{\tau_0}{t_{pulse}}\right), \tag{3}$$

where $I_{sw,0}$ is the intrinsic critical switching current and $\tau_0$ the precession time. Through eqn. (3), we extract $\tau_0 \approx 0.06$ ns for type-*x* geometry and $\tau_0 \approx 1.11$ ns for type-*y*. Therefore, type-*x* indeed has a potential in high speed SOT switching applications but only in the sub-ns regime.

### e. Canting angle dependence of SOT switching

Previous studies have demonstrated that type-*x* devices can achieve field-free SOT switching by tilting orientation of the FM pillar major axis towards *y* direction to break symmetry [16,19,31]. In this section, we further simulate the pillar canting angle dependence of $J_{sw}$. The schematics of the FM pillar for type-*x* being canted from the current channel direction by $\varphi = 1$ to 179 deg are shown in Fig. 3(c). Note that no external fields are included in the simulations in this section. As shown in Fig. 3(d), field-free switching can be achieved with $J_{sw} = 2.3 \times 10^{12}$ A/m$^2$ for the type-*x* geometry at $\varphi = 1$ deg. The trend of $J_{sw}$ vs. $\varphi$ is almost symmetric about 90 deg, as expected. When the canting angle $\varphi$ increases from 1 deg to 90 deg, the switching current density decreases rapidly at $\varphi \sim 21$ deg, which indicates that the magnetization dynamics transform from type-*x* to type-*y* behavior at this particular canting angle. We can therefore divide the angle dependence into two zones: Those with higher $J_{sw}$ ($\varphi = 1$ deg to 21 deg, 159 deg to 179 deg) belong to the 'type-*x* regime'(the yellow region in Fig.3 (d)), while the other part ($\varphi = 22$ deg to 158 deg) with a much smaller $J_{sw}$ is the 'type-*y* regime' (the orange region in Fig.3 (d)).

To verify the canting angle effect from macrospin simulations, we prepare a series of W(3)/CoFeB(2.5)/MgO(1) (numbers in parentheses are in nanometers) Hall-bar devices with lateral dimensions of 5×60 µm$^2$, as schematically shown in Fig. 4(a). The details of thin film and device preparation can be found in a previous study [14]. These devices are further annealed at 220°C for 20 mins with applying an in-plane magnetic field of 0.25 T along a specific direction with respect to the Hall-bar device to establish easy axis. The actual orientation of the easy axis with respect to the current channel in these devices ($\varphi_{EA}$) are further determined by angle dependent anisotropic magnetoresistance (AMR) measurements (details of the determination of $|\varphi_{EA}|$ can be found in Supplementary Materials S2 [39]). Through the detection of spin Hall effective field-modified AMR or unidirectional MR (UMR)[14,43,44], current-induced magnetization switching can be observed in these samples under zero-field condition. When the SOT-induced switching occurs, a $\Delta R_{xx}$ can be detected [14]. As shown in Fig. 4(b), the switching current $I_c$ measured with $t_{pulse}$ = 50 ms decreases from 6.0 mA to 2.3 mA with increasing $\varphi_{EA}$.

We further perform pulse width dependent SOT switching measurement. As shown in Fig. 4(c), the SOT-induced magnetization switching for such long pulse width (50 ms ≤ $t_{pulse}$ ≤ 1 s) is a thermally activated process and the critical switching current $I_c$ becomes larger as $t_{pulse}$ decreases. The zero-thermal fluctuation critical switching current $I_{c0}$ can be further obtained by [45]:

$$I_c = I_{c0}\left(1 - \frac{1}{\Delta}\ln\left(\frac{t_{pulse}}{\tau_a}\right)\right), \qquad (4)$$

where $\Delta$ is the thermal stability factor and $\tau_a \approx$ 1 ns is the attempt time scale for thermally activated switching.[46] By using the data in Fig. 4(c) and eqn. (4), we find that $J_{c0} \approx 4.63 \times 10^{11}$ A/m² with $\Delta \approx$ 64 for the $|\varphi_{EA}|$ = 1.2 deg device and $J_{c0} \approx 0.99 \times 10^{11}$ A/m² with $\Delta \approx$ 30 for the $\varphi_{EA}$ = 80.3 deg device. $J_{c0}$ for all annealed Hall-bar devices with different $\varphi_{EA}$ are summarized in Fig. 4(d). The zero-thermal fluctuation critical switching current density sharply decreases from $J_{c0} \approx 4.63 \times 10^{11}$ A/m² for $|\varphi_{EA}|$ = 1.2 deg to $J_{c0} \approx 1.19 \times 10^{11}$ A/m² for $|\varphi_{EA}|$ = 33.9 deg and almost saturated at $J_{c0} \approx 1.00 \times 10^{11}$ A/m² for $|\varphi_{EA}|$ = 80.3 deg. The trend of $J_{c0}$ vs.

|$\varphi_{EA}$| from the experiments is qualitatively consistent with that from the simulations, which suggests that the canting angle effect on switching current density exists even in micro-sized devices. The experimental result shows a more gradual trend and a larger threshold canting angle compared to Fig. 3(d), which is attributed to the existence of the FLT in our samples. Note that a similar trend between $J_c$ and $\varphi_{EA}$ has also been observed in the in-plane magnetized elliptic nanodot array with various easy axis directions through differential planar hall effect measurement [31].

**f. Field-like torque contribution and the switching phase diagram**

In most of the previous SOT switching studies, DLT has been regarded as the primary driving mechanism. Recent theoretical studies suggest the addition FLT with an appropriate magnitude can assist magnetization switching and apparently reduce the switching current [47,48]. This sizeable FLT, which is a combined effect from the bulk SHE, interfacial REE, and Oersted field, has been observed in various HM/FM heterostructures via spin torque ferromagnetic resonance (ST-FMR) measurement and harmonic measurement [49-51]. It is therefore valid to say that the possible contribution of FLT in SOT switching cannot be entirely neglected.

To systematically investigate the effect of FLT in a type-$x$ system, we introduce an additional FLT term in eqn. (1), which is expressed as

$$\tau_{FL} = \gamma \frac{\hbar(\eta \xi_{DL})J}{2eM_s t_{FM}} (\boldsymbol{\sigma} \times \mathbf{m}), \tag{4}$$

where $\eta$ is a dimensionless FLT/DLT ratio $\equiv \tau_{FL}/\tau_{DL}$. In this section, pillar orientation of the type-$x$ structure is canted as 1 deg to achieve field-free switching. The applied current density is swept from 0.0 to 11.0 ×$10^{12}$ A/$m^2$ with a 5 ns duration and we record the *x*-component of magnetization after a 5-ns-relaxation. By varying the FLT/DLT ratio from +1 to -1 (corresponds to $\xi_{FL}$ = - 0.5 to + 0.5), we arrive at the phase diagram shown in Fig. 5(a), in which the white (red) region in the phase diagram represents the magnetization being switched (not being switched). This phase diagram can be divided into four regions, namely non-switching, random switching, toggle switching and deterministic switching region.

For the part where FLT/DLT ratio > 0 in the switching phase diagram [Fig. 5(a)], it contains a deterministic switching region (white) and a non-switching region (red). The deterministic switching region expands with increasing FLT/DLT ratio because FLT of the same sign can accelerate energy dissipation and facilitate the magnetization stabilization. In this region, a larger FLT can significantly reduce the critical switching current density.

To further gain insight into the switching dynamics for FLT/DLT ratio < 0, we extract magnetization trajectories in the non-switching region, random switching region, toggle switching region and deterministic switching region in the phase diagram for FLT/DLT ratio = -0.4. When the applied current density is large enough, deterministic switching can be observed. As shown in Fig. 5(b), the magnetization is aligned along +*y* but with a slight deviation (towards *x* direction) from it due to the small canting angle. After relaxation, the magnetization can be successfully switched. In the deterministic switching region, DLT is the dominating term to drive magnetization switching.

As the current density is lowered down to ~ 8.42×$10^{12}$ A/$m^2$, toggle switching will occur. As shown in Fig. 5(c), the magnetization will stop at a position with a large +*z* component before relaxation. For lower current densities, the magnetization before relaxation will stabilize at a position with a higher +*z* component, smaller +*y* and +*x* components. The magnitude of $m_z$ after the excitation of SOT will affect the final orientation of magnetization (switched or not switched) after relaxation, since different $m_z$ will lead to different precession cycle number. We can define the precession number $N$ as the number of times the magnetization passes $y$-$z$ plane during the relaxation process. In this region, a smaller applied current density corresponds to larger $m_z$ and a larger $N$. As current density decreases, the behavior of $N$ becomes a staircase function. If $N$ is an odd (even) number, we define the normalized $N$, $N_{nor}$, as +1 (-1). As shown in Fig. 5(d), depending on $N$, magnetization switches after half cycles of precession ($N_{nor}$ = -1) and does not switch after full cycles of precession ($N_{nor}$ = +1). This result is also consistent with an earlier macrospin study that considered a non-zero FLT/DLT ratio in the type-$z$ system [48].

For the random switching region, there exists two types of magnetization dynamics during the excitation of SOT, namely precession and oscillation. The magnetization trajectories of precession and oscillation dynamics are shown in Fig. 5(e) and Fig. 5(f), respectively. This kind of steady tilted precession is caused by the dynamical equilibrium between the FLT, DLT and demagnetization field torque with a larger applied current density. The final state of magnetization also depends on the precession number $N$ during the 5-ns relaxation. For a lower applied current density in the random switching region, the oscillation dynamics emerges due to the competition between the demagnetization field torque and the SOT. The sign of the $x$ component of the final magnetization position before relaxation decides whether the magnetization is switched or not. If the magnetization has a -$x$ component before relaxation, the magnetization will precess and stop at an almost -$x$ direction, which means the magnetization switching is achieved. The existence of precession and oscillation modes has also been reported in a previous study [52], which points out that the magnitude of the FLT/DLT ratio can be modified by tuning the HM and FM layer thicknesses.

For the non-switching region, a representative magnetization trajectory is shown in Fig. 5(g). The DLT is too small to overcome the energy barrier in the *y-z* plane before relaxation, therefore the magnetization will stabilize to the initial position after relaxation. In short, FLT plays a crucial role in type-*x* SOT-driven magnetization switching. For FLT/DLT > 0, the FLT can assist the magnetization switching and reduce the critical switching current density robustly. For FLT/DLT < 0, precession dominates the magnetization dynamics, thereby suppressing the magnetization switching. This feature might be disadvantageous for memory applications but could be useful in probabilistic computing related applications [53-55]. In the following section, we will focus on the potential of adopting type-*x* SOT switching with FLT/DLT > 0.

### g. Field-like torque assisted type-*x* switching

Next, we study the variation of switching current density with respect to the external field $\mu_0 H_z$ and the canting angle $\varphi$ with the inclusion of different FLT/DLT ratios. $J_{sw}$ vs. $\mu_0 H_z$ with various FLT/DLT ratios are shown in Fig. 6(a) ($\varphi = 0$ deg). As the FLT/DLT ratio increases, the slope of $J_{sw}$ vs. $\mu_0 H_z$ decreases significantly due to the enhancement of FLT effective field. However, the *x*-intercept remains constant ($\mu_0 H_{z,0} = 0.40$ T) for each FLT/DLT ratio. The canting angle dependences of $J_{sw}$ at various FLT/DLT ratios are shown in Fig. 6(b), where the overall $J_{sw}$ becomes lower with increasing FLT. The type-*x* regime also expands toward much closer to 90 deg due to the faster energy dissipation during precession as assisted by the positive FLT.

To further investigate variations between type-*x* ($\varphi = 1$ deg) and type-*y* ($\varphi = 90$ deg) system at short current pulse width regime, we calculate the switching current $I_{sw}$ for these two types of systems with $0.1$ ns $\leq t_{pulse} \leq 10$ ns and a 5-ns relaxation at various FLT/DLT ratios. The current channel width and the shape of the pillar are set to be the same as Fig. 4(c) for type-*x* ($\varphi = 1$ deg) and type-*y* ($\varphi$ = 90 deg). The $I_{sw}$ vs. $t_{pulse}$ curve for FLT/DLT ratio = 0.4 is shown in Fig. 6(c). Notice that the crossover point ($t_{CO}$) between type-*x* and type-*y* systems migrates from 0.23 ns for FLT/DLT = 0.0 to 2.66 ns for FLT/DLT = 0.4. $t_{CO}$ for each FLT/DLT ratio are summarized in Fig. 6(d), in which the gray region represents that $I_{sw}$ of type-*y* is larger than that of type-*x* (more advantageous to adopt type-*x*).

Again, the intrinsic critical switching current $I_{sw,0}$ and the precession time $\tau_0$ can be obtained from $I_{sw,0}$ vs. $t_{pulse}$ curves for each FLT/DLT ratio through eqn. (3). Figures 6(e) and 6(f) show $I_{sw,0}$ and $\tau_0$ vs. FLT/DLT ratio. $I_{sw,0}$ of type-*x* structure decreases with increasing the FLT/DLT ratio and becomes smaller than that of type-*y* at FLT/DLT = 0.5. For type-*y*, the precession time $\tau_0$ decreases from 1.8 ns to 0.3 ns as increasing the FLT/DLT ratio from 0.0 to 1.0. For type-*x*, $\tau_0$ also drops from 50 ps to 0.4 ps as the FLT/DLT ratio increases. However, type-*x* has a much smaller $\tau_0$ for the whole range of the FLT/DLT ratio, indicative the potential of a much faster switching with type-*x* structure. Overall, we show that the performance of type-*x* (lower $I_{sw}$ and lower $\tau_0$) can be better than that of the type-*y* system with an appropriate FLT in the short pulse regime.

Note that to achieve a sizable FLT to DLT ratio, there exists several possible feasible ways. One approach is to select a suitable spin Hall metal which intrinsically possesses large positive FLT to DLT ratio such as Ta [56,57] or Hf [58]. Second, interfacial engineering is a method to increase the positive FLT to DLT ratio as well. It is shown that the ratio is enhanced via the insertion of a light metal layer between the spin Hall metal and the FM layer [59]. Moreover, annealing temperature can also be optimized to obtain a large FLT to DLT ratio [60].

## h. Micromagnetic simulations

To mimic magnetic dynamics in real devices with type-$x$ structure, we further perform micromagnetic simulations with Mumax3 [61]. Additional materials parameters included in micromagnetic simulations are summarized in Table. 1. The exchange stiffness constant is set as $1.6 \times 10^{-11}$ J/m and the strength of interfacial Dzyaloshinskii-Moriya interaction (DMI) is $2 \times 10^{-4}$ J/m$^2$, which are both referred to the typical values for CoFeB [62]. The macrospin simulations show that $K_u$ does not prominently influence $J_{sw}$. It remains at the same order even when $K_u$ is added up to $10^5$ J/m$^3$ with the same $\mu_0 H_z$ applied. Consequently, we tentatively set $K_u$ as zero to conduct micromagnetic simulations for simplicity.

Similar to the macrospin result, Fig. 7(a) shows that $J_{sw}$ decreases linearly as $\mu_0 H_z$ increases until it reaches zero at around $\mu_0 H_z = 0.39$ T. The slope fitted is $-5.8 \times 10^{12}$ A/Tm$^2$, which is also in line with the macrospin result (Fig. 2(a)). Fig. 7 (b) shows $J_{sw}$ as a function of $\varphi$ from 3 deg to 177 deg for various FLT/DLT ratios. For small FLT/DLT ratios (0 and 0.2), the transitions from a type-$x$-like to type-$y$-like behavior appear at $\varphi \sim 27$ deg and 153 deg, which are slightly different from the results of macrospin simulations. This slight discrepancy is tentatively attributed to the inclusion of DMI and exchange interactions in micromagnetic simulations. At either type-$x$ or type-$y$ regime, $J_{sw}$ is smaller as the canting angle approaches 90 deg (*i.e.* $J_{sw}$ decreases as the system looks more analogous to type-$y$), meaning that type-$x$ is not as efficient as type-$y$ with a small FLT. However, $J_{sw}$ drops significantly and the type-$x$ regime broadens as FLT/DLT ratio becomes larger. When FLT/DLT ratio = 1, the boundary between type-$x$ and type-$y$ regime becomes ambiguous. Again we observe that type-$x$ is inferior to type-$y$ when FLT is small or absent but may become advantageous over type-$y$ when the FLT is large enough.

Next, we look into the performance of type-$x$ ($\varphi = 1$ deg) and type-$y$ ($\varphi = 90$ deg) structures when short current pulses ($t_{pulse} \leq 5$ ns) are applied. Fig. 7(c) shows critical switching current $I_{sw}$ as a function of current pulse width $t_{pulse}$ for various positive FLT/DLT ratios. Note that the relaxation time is 5 ns and the width of the current channel considered here is the same as that shown in Fig. 3(c). The data points are all fitted with eqn. (3). The extracted $I_{sw,0}$ and $\tau_0$ from each curve and $t_{CO}$ are summarized in Table 3. For current pulses shorter than $t_{CO}$, type-x ($\varphi = 1$ deg) has smaller $I_{sw}$ than type-y, indicating that type-x has better performance within this regime. As FLT/DLT ratio increases from 0 to 0.4, $t_{CO}$ increases from 0.60 to 3.13 ns. The trend of $t_{CO}$ vs. FLT/DLT ratio again suggests that from the perspective of designing type-x SOT-MRAM, sizable FLT can be beneficial in the reduction of $I_{sw}$ and the enhancement of $t_{CO}$.

Finally, we simulate the FLT/DLT ratio dependence of $I_{sw}$. Note that the pulse duration and relaxation time here are both fixed as 5 ns. Fig. 7(d) shows $I_{sw}$ as a function of positive FLT/DLT ratio for type-x ($\varphi = 1$ deg) and type-y structures. As shown in Fig. 7(d), $I_{sw}$ of type-y only decreases slightly with increasing FLT/DLT ratio, while that of type-x is inversely proportional to FLT/DLT ratio. The results show that when FLT/DLT ratio is larger than 0.5, type-x has smaller $I_{sw}$ than type-y, which is consistent with Fig. 6(e) obtained by macrospin simulation. This trend again suggests that FLT indeed plays a critical role in promoting the application of type-x devices.

### III. CONCLUSION

In summary, we scrutinize the features of type-x SOT switching scheme in nano-sized HM/FM bilayer structures by numerical simulations. Through macrospin simulations, we demonstrate that the switching current density $J_{sw}$ depends on $\mu_0 H_z$ and the shape anisotropy field (including $\mu_0 H_{k,in}$ and $\mu_0 H_{k,out}$). We show that type-x is more efficient than type-y only when the applied pulse is shorter than a crossover timescale $t_{CO} \sim 0.23$ ns. The trend of $\varphi$ vs. $J_{sw}$ obtained by simulation is consistent with the experimental results. To explore the effects of FLT, we obtain a detailed switching phase diagram that provides insight into the magnetization dynamics for applications using type-*x* geometry. Moreover, by both macrospin and micromagnetic simulations, we show that a positive FLT/DLT ratio reduces switching current $I_{sw}$ considerably and the window of ultrafast type-*x* SOT switching can be expanded with increasing FLT. Therefore, by optimizing the geometry of the FM pillar and adopting materials with suitable FLT efficiency, type-*x* geometry is anticipated to be a promising scheme for future SOT-MRAM applications.

## ACKNOWLEDGEMENT

This work is supported by the Ministry of Science and Technology of Taiwan (MOST) under grant No. MOST-110-2636-M-002-013 and by the Center of Atomic Initiative for New Materials (AI-Mat) and the Advanced Research Center of Green Materials Science and Technology, National Taiwan University from the Featured Areas Research Center Program within the framework of the Higher Education Sprout Project by the Ministry of Education (MOE) in Taiwan under grant No. NTU-110L9008. We also thank Shy-Jay Lin of Taiwan Semiconductor Manufacturing Company for fruitful discussions.

Table 1. The switching characteristics of type-z, type-y and type-x geometries.

|  | Type-z | Type-y | Type-x |
| --- | --- | --- | --- |
| Magnetic anisotropy | PMA | IMA | IMA |
| Easy axis (EA) direction | EA ⊥ x-y plane<br>EA ⊥ $\hat{\sigma}$ | EA // y<br>EA // $\hat{\sigma}$ | EA // x<br>EA ⊥ $\hat{\sigma}$ |
| Achieving field-free switching? | No, a $\mu_0 H_x$ is required. | Yes | No, a $\mu_0 H_z$ is required. |
| The methods to achieving field-free switching | 1. An additional voltage-controlled electric field [21]<br>2. Exchange bias field [22]<br>3. Wedge structure [12,23]<br>4. An additional FM layer with IMA [24-26]<br>5. Geometrical domain-wall pinning [27]<br>6. Competing spins [28,29]<br>7. Localized laser annealing [30] | No need | 1. Canted EA in FM layer [16,19,31]<br>2. An additional FM layer with PMA [16] |

Table 2. Detailed parameters adopted in macrospin and micromagnetic simulations.

|  | Symbol (unit) | Macrospin simulation | Micromagnetic simulation |
| --- | --- | --- | --- |
| Saturation magnetization | $M_s$ (emu/cc) | 1100 | 1100 |
| Damping constant | $\alpha$ | 0.02 (for IMA)<br>0.1 (for PMA) | 0.02 |
| Uniaxial anisotropy energy density | $K_u$ (J/m$^3$) | $-8 \times 10^5$ (for IMA)<br>$+8 \times 10^5$ (for PMA) | 0 |
| FM pillar size | $V_{pillar}$ (nm$^3$) | 60×30×1 | 60×30×1 |
| Thickness (HM layer) | $t_{HM}$ (nm) | 3 | 3 |
| DL-SOT efficiency | $|\xi_{DL}|$ | 0.5 | 0.5 |
| Interfacial DMI energy density | $D$ (J/m$^2$) | – | $2\times10^{-4}$ |
| Exchange stiffness constant | $A$ (J/m) | – | $1.6\times10^{-11}$ |
| Cell size | $V_{cell}$ (nm$^3$) | – | 3×3×1 |

Table 3. Fitted parameters ($I_{sw,0}$ and $\tau_0$) of type-*y* and type-*x* switching with different FLT/DLT ratios according to eqn. (3). $t_{CO}$ is the crossover point of the type-*x* curve with respect to the type-*y* curve.

|  | $I_{sw,0}(\mu A)$ | $\tau_0$(ns) | $t_{CO}$(ns) |
|---|---|---|---|
| Type-y ($\varphi$ = 90 deg) FLT/DLT = 0.0 | 19.71 | 4.00 | NA |
| Type-x ($\varphi$ = 1 deg) FLT/DLT = 0.0 | 128.16 | 0.11 | 0.60 |
| Type-x ($\varphi$ = 1 deg) FLT/DLT = 0.2 | 77.90 | 0.051 | 1.29 |
| Type-x ($\varphi$ = 1 deg) FLT/DLT = 0.4 | 44.85 | 0.0033 | 3.13 |

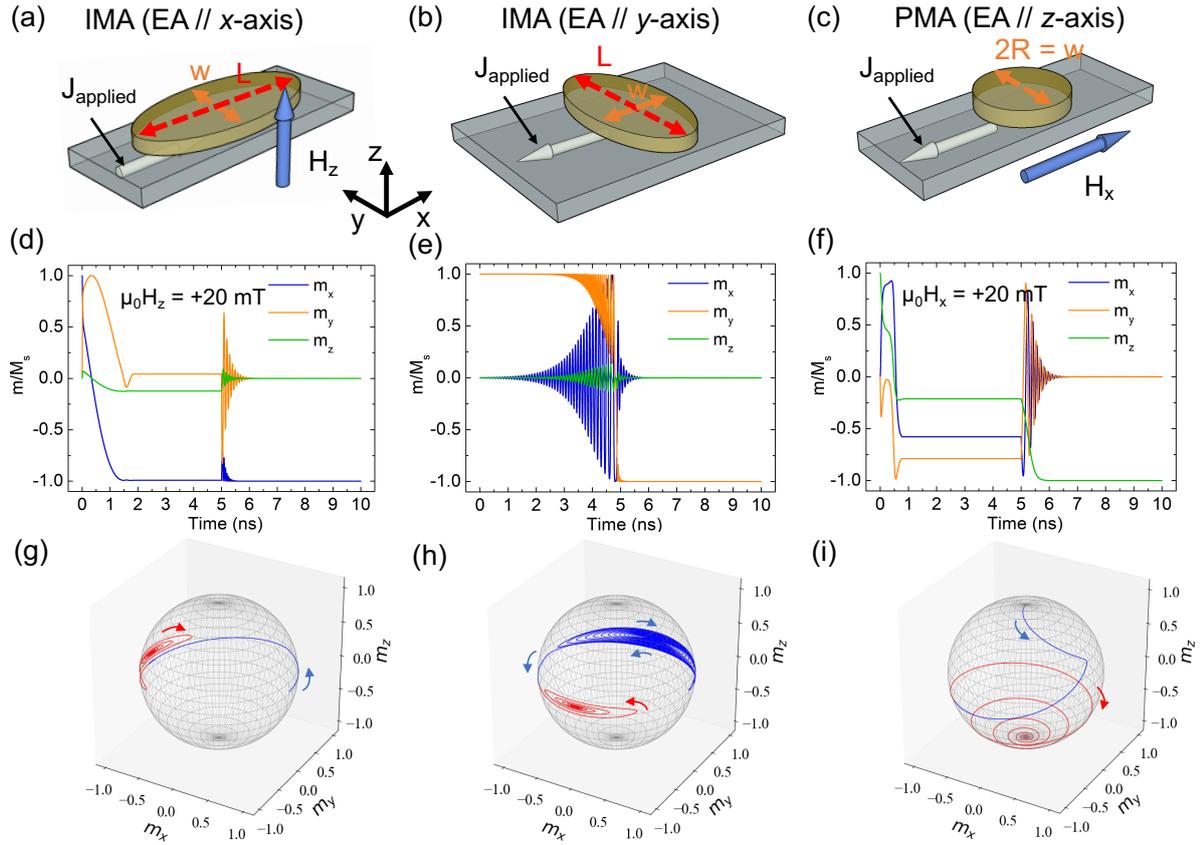

Figure 1. Macrospin simulations: Schematic illustrations of (a) type-*x*, (b) type-*y* and (c) type-*z* SOT switching. IMA and PMA represent that FM pillar has in-plane magnetic anisotropy and out-of-plane magnetic anisotropy, respectively. The magnetization dynamics of (d) type-*x*, (e) type-*y* and (f) type-*z* schemes under the excitation of SOT for 5 ns. 3D magnetization trajectories of (g) type-*x*, (h) type-*y* and (i) type-*z* switching. The external field (20 mT) along +*z* direction and +*x* direction is required to achieve deterministic switching for type-*x* and type-*z* structure, respectively. In (d, e, f), the *x*, *y* and *z* components of magnetization $m_x$, $m_y$ and $m_z$ are denoted by the blue, orange and green lines, respectively. In (g, h, i), the trajectories are divided into excitation segments (blue lines) and relaxation segments (red lines).

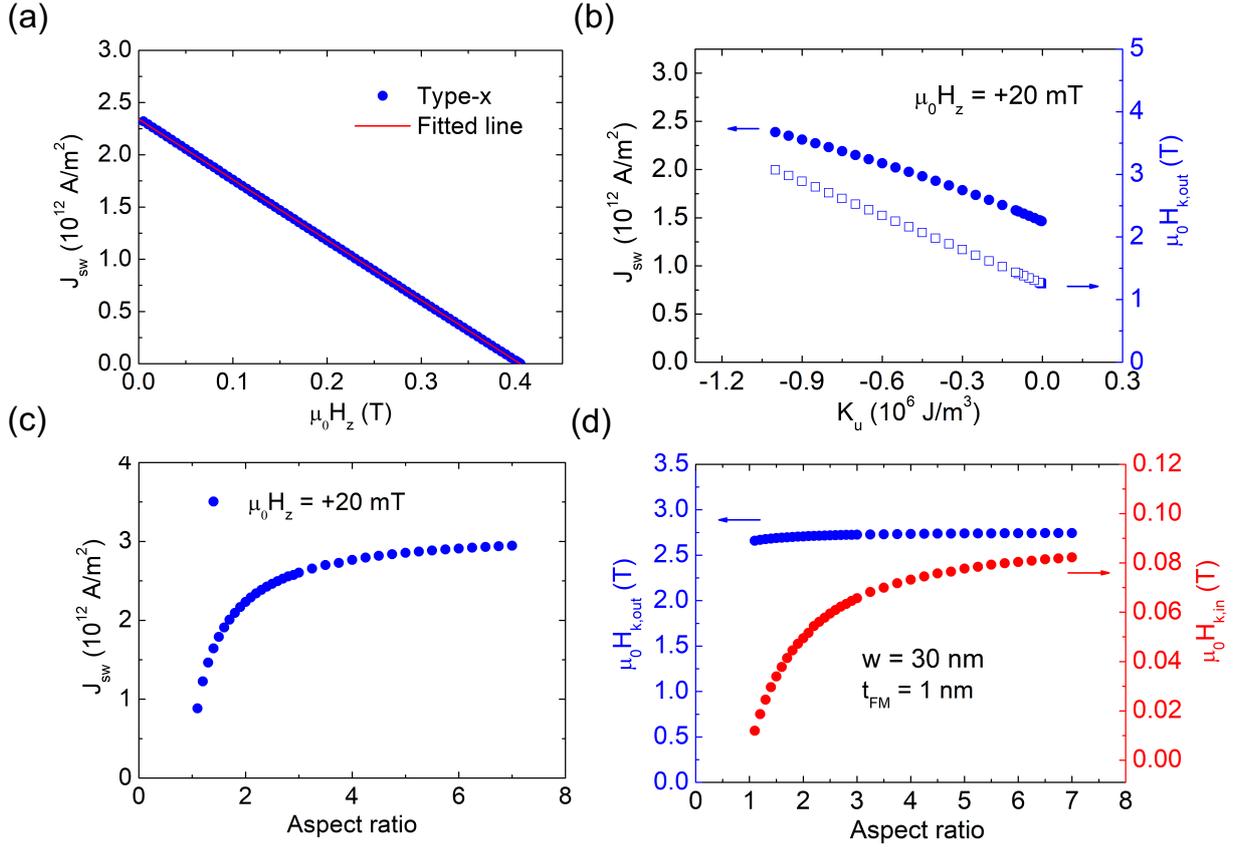

Figure 2. Macrospin simulations: Critical switching current density $J_{sw}$ for type-$x$ geometry as a function of (a) $z$ direction external field $\mu_0H_z$ and (b) magnetic anisotropy energy density $K_u$. The representative device geometry is set as $60 \times 30 \times 1$ nm$^3$, $K_u$ is $-8 \times 10^5$ J/m$^3$ with an uniaxial anisotropy unit vector $u = [0, 0, 1]$. The open squares in (b) represent the magnitude of $\mu_0H_{k,out}$. The dots in (b) represent the magnitude of $J_{sw}$. (c) $J_{sw}$, (d) $\mu_0H_{k,out}$ and $\mu_0H_{k,in}$ for type-$x$ structure as functions of the FM pillar aspect ratio.

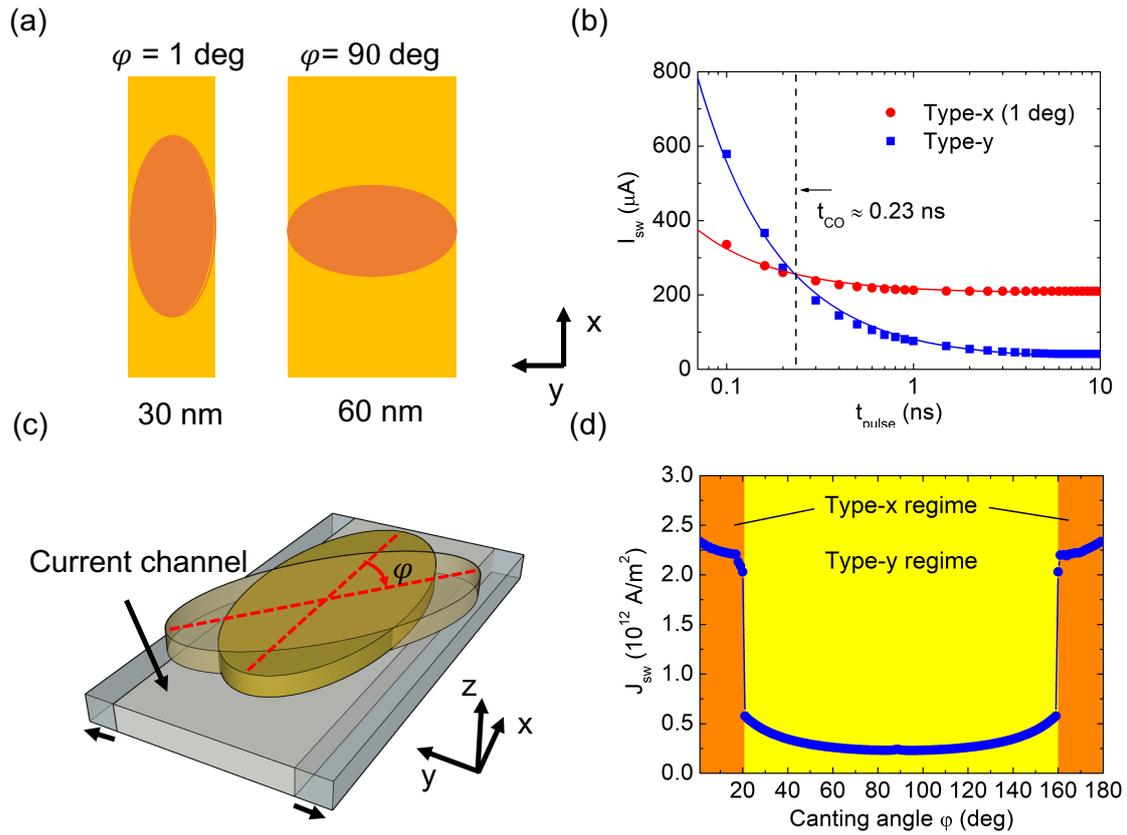

Figure 3. Macrospin simulations: (a) Schematics of the simulated device with pillar canting angle $\varphi = 1$ deg (type-$x$) and $\varphi = 90$ deg (type-$y$). (b) Current pulse width $t_{\text{pulse}}$ dependence of switching current $I_{\text{sw}}$ for type-$x$ ($\varphi = 1$ deg) and type-$y$ ($\varphi = 90$ deg) structure. The red and black lines are fitted by eqn. (3). The dashed line represents the crossover point $t_{\text{CO}}$ between the type-$x$ and type-$y$ geometry. (c) Schematics of the simulated device being canted away from the $x$ direction. (d) $J_{\text{sw}}$ calculated by varying $\varphi$ under zero external magnetic field.

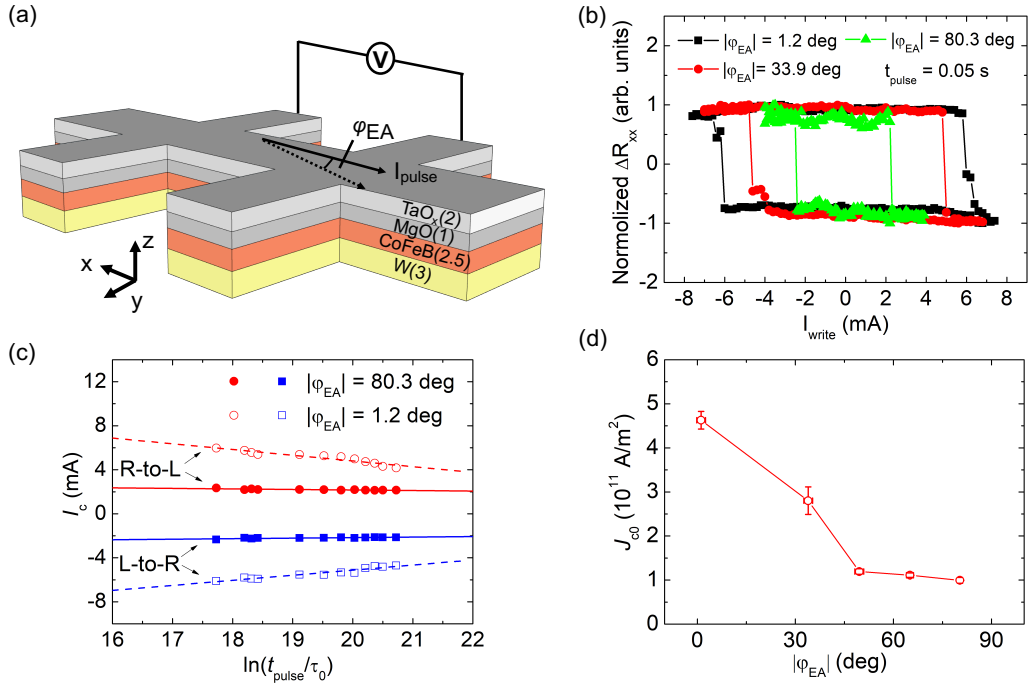

Figure 4. Experiments: (a) Schematic illustration of a W(3)/CoFeB(2.5)/MgO(1) Hall-bar device and current-induced magnetization switching measurement. The direction of the positive current is defined along -x direction. (b) Representative $\Delta R_{xx}$ as a function of write current $I_{write}$ for W(3)/CoFeB(2.5)/MgO(1) devices with the different $\varphi_{EA}$. (c) The write current pulse width dependence of the critical switching current $I_c$ for W(3)/CoFeB(2.5)/MgO(1) devices. (d) The zero-thermal fluctuation critical switching current density $J_{c0}$ as a function of the easy axis canting angle $\varphi_{EA}$.

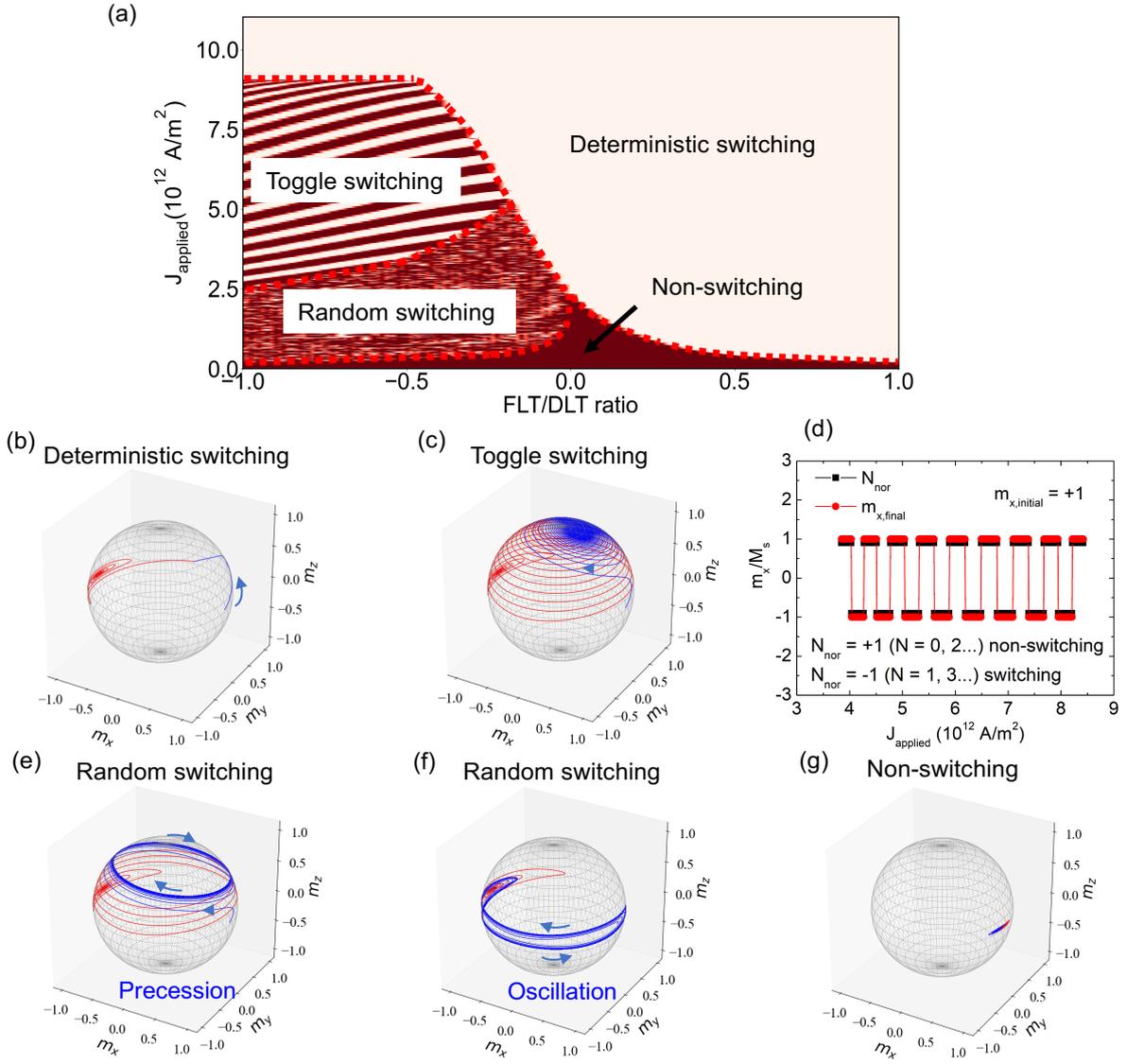

Figure 5. Macrospin simulations: (a) The switching phase diagram of a type-$x$ device ($\varphi = 1$ deg) at various FLT/DLT ratios with a fixed DLT efficiency = 0.5. Magnetization trajectories for FLT/DLT = -0.4 in the (b) deterministic switching region with $J_{applied} = 9.56 \times 10^{12}$ A/m$^2$, (c) toggle switching region with $J_{applied} = 4.19 \times 10^{12}$ A/m$^2$, (e) precession dynamics before relaxation in the random switching region with $J_{applied} = 2.21 \times 10^{12}$ A/m$^2$, (f) oscillation dynamics before relaxation in the random switching region with $J_{applied} = 0.37 \times 10^{12}$ A/m$^2$, and (g) non-switching region with $J_{applied} = 0.15 \times 10^{12}$ A/m$^2$. (d) The normalized precession number $N_{nor}$ and $m_x$ as functions of $J_{applied}$ in the toggle switching region with FLT/DLT ratio = -0.4.

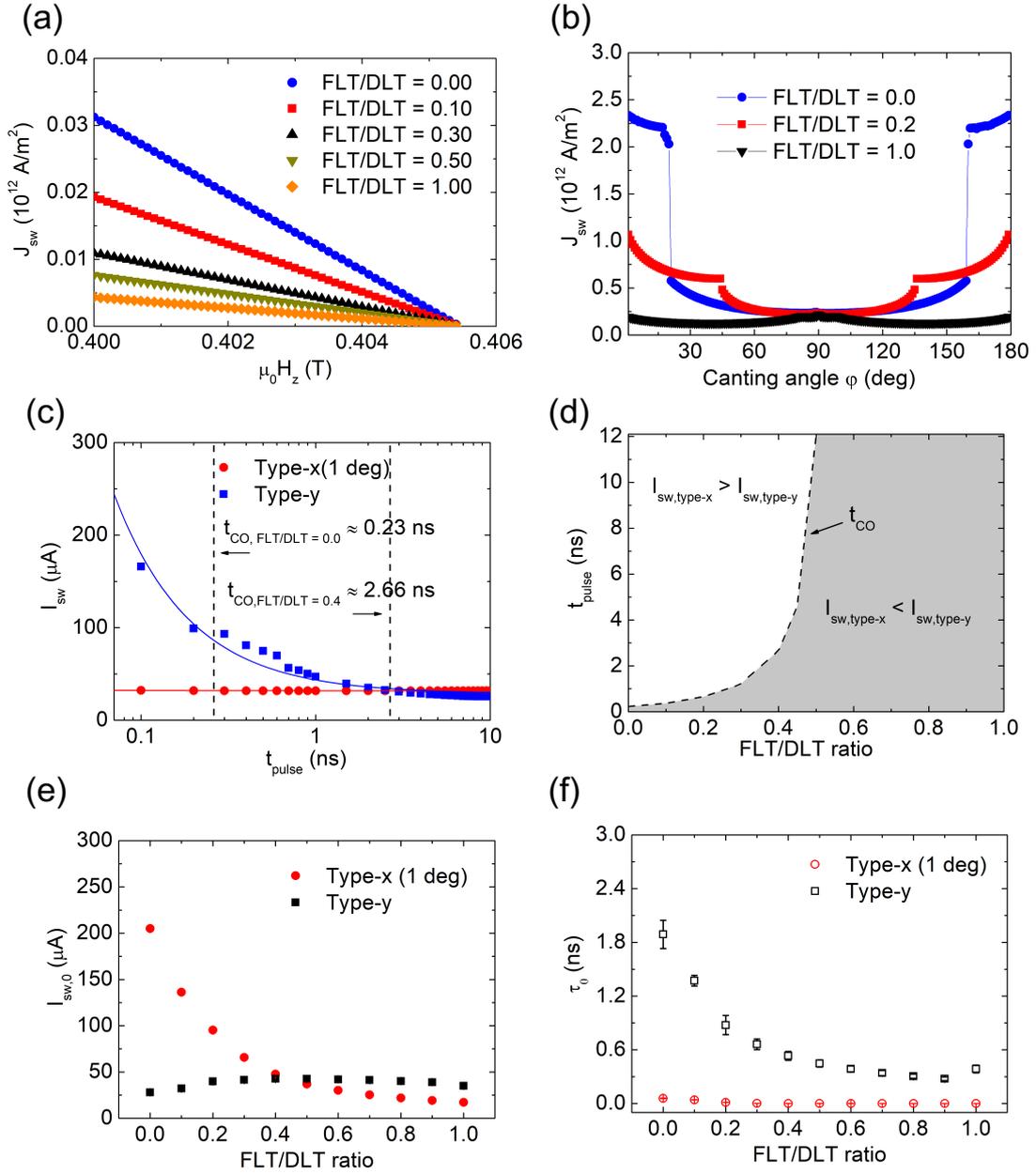

Figure 6. Macrospin simulations: (a) $J_{sw}$ vs. $\mu_0 H_z$ for type-$x$ structure with various FLT/DLT ratios. (b) $J_{sw}$ vs. canting angle ($\varphi$ = 1 to 179 deg) with various FLT/DLT ratios. (c) $t_{pulse}$ dependence of $J_{sw}$ for type-$x$ ($\varphi$ = 1 deg) and type-$y$ ($\varphi$ = 90 deg) structure with FLT/DLT ratio = 0.4. The dashed lines in (c) represent the crossover point $t_{CO}$ between the type-$x$ and type-$y$ geometries. (d) $I_{sw}$ phase diagram for type-$x$ ($\varphi$ = 1 deg) and type-$y$ ($\varphi$ = 90 deg) structures. The black dashed line in (d) represents $t_{CO}$ between the type-$x$ and type-$y$ geometries with various positive FLT/DLT ratios. (e) $I_{sw,0}$ and (f) $\tau_0$ for type-$x$ ($\varphi$ = 1 deg) and type-$y$ ($\varphi$ = 90 deg) structures extracted by eqn. (3) as functions of the FLT/DLT ratio.

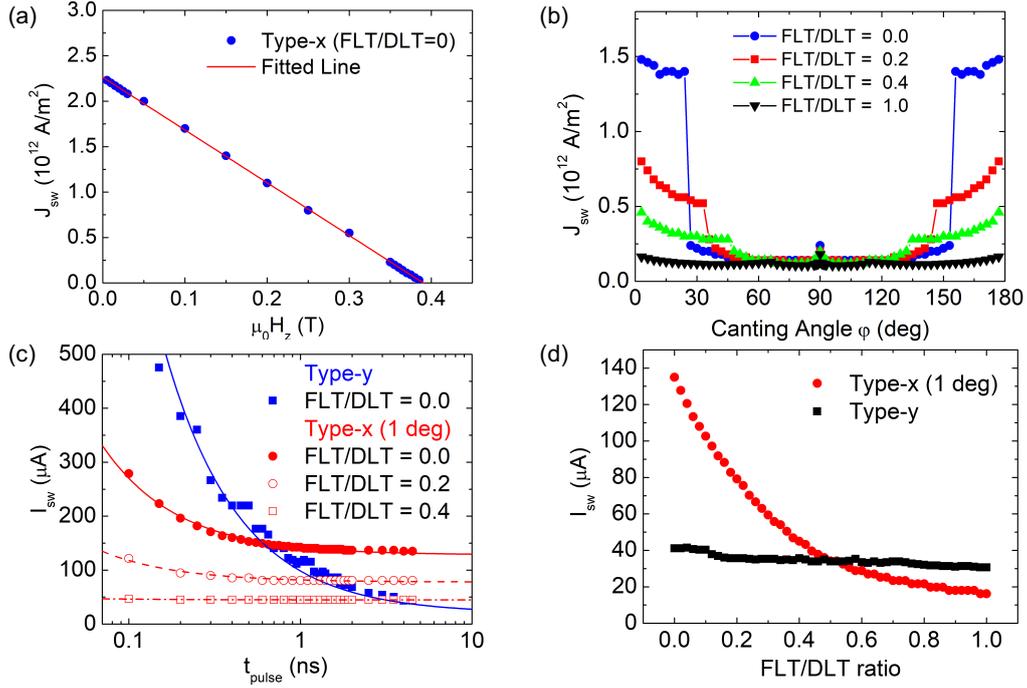

Figure 7. Micromagnetic simulations: (a) $\mu_0 H_z$ dependence of $J_{sw}$ for type-$x$ structure obtained by micromagnetic simulations. (b) Canting angle dependence of $J_{sw}$ with various FLT/DLT ratios. (c) Pulse width dependence of $I_{sw}$ with various FLT/DLT ratios and (d) FLT/DLT ratio dependence of $I_{sw}$ for type-$x$ ($\varphi = 1$ deg) and type-$y$ ($\varphi = 90$ deg) structures. Note that in (c) the type-$y$ curve is simulated without FLT and other type-$x$ curves are simulated with $\varphi = 1$ deg.